\documentclass[aps,prl,showpacs,twocolumn,amsmath,amssymb,superscriptaddress]{revtex4-1}
\usepackage{graphicx}
\usepackage{dcolumn}
\usepackage{bm}
\usepackage{units}
\usepackage{upgreek}
\usepackage{ucs}
\usepackage{color}
\usepackage{cancel}
\usepackage{hyperref}
\usepackage[all]{hypcap}
\usepackage{braket}

\begin{document}

\title{Chaotic level mixing in a two-band Bose-Hubbard model}
\author{Carlos A. Parra-Murillo}
\affiliation{Departamento de F\'isica - ICEX - UFMG Cx. Postal. 702, 30.161-970, Belo Horizonte, MG, Brasil}
\author{Javier Madro\~nero}
\affiliation{Departamento de F\'isica, Universidad del Valle, Cali, Colombia}
\author{Sandro Wimberger}
\affiliation{Dipartimento di Fisica e Science della Terra, Universit\`{a} di Parma Via G.P. Usberti 7/a, 43124 Parma, Italy}
\affiliation{INFN, Sezione di Milano Bicocca, Gruppo Collegato di Parma, Italy}
\email{sandromarcel.wimberger@unipr.it}

\begin{abstract}
We present a two-band Bose-Hubbard model which is shown to be minimal in the necessary coupling terms at resonant tunneling conditions. The dynamics of the many-body problem is studied by sweeping the system across an avoided level crossing. The linear sweep generalizes Landau-Zener transitions from single-particle to many-body realizations. The temporal evolution of single- and two-body observables along the sweeps is investigated in order to characterize the non-equilibrium dynamics in our complex quantum system.
\end{abstract}

\maketitle
\section{Introduction}\label{sec:1}

Describing stationary ground states and transport dynamics in real solid-states is notoriously difficult because of the too many degrees of freedom involved. It is the merit of simplified models, which nevertheless take into account the relevant physical mechanisms, to allow analytical and numerical predictions for such complex quantum systems. A celebrated model is the Hubbard model first introduced for fermions, whose main approximations are that is it discrete describing a lattice system and that interactions are typically short ranged \cite{FermiH}. 

The realization of lattice models with ultracold atoms has given new impetus to the theoretical study of many-body models, be it for fermions or bosons. Neglecting the spin degree of freedom, the simplest Hubbard model in one spatial dimension is integrable for fermions, while its counterpart for bosons is shown to be non-integrable \cite{Kolo2004}. This originates in the fact that too many combinations exist of distributing bosons on a lattice. The Bose-Hubbard model we are referring to now and in the following is given by the many-body Hamiltonian \cite{BDZ2008}
\begin{eqnarray}\label{eq:1}
 \hat{H}_{\rm BHM} &=&\sum_{l}\left[-\frac{J}{2}\left(\hat{a}^{\dagger}_{l+1}\hat{a}_{l}+h.c.\right) + \frac{W}{2}\hat{a}^{\dagger 2}_{l}\hat{a}^2_{l}\right] \;.
 \end{eqnarray}
Here $\hat{a}_l$ and $\hat{a}_l^\dagger$ are the bosonic annihilation and creation operators in the lattice mode $l$, $W$ denotes the interaction strength and $J$ the tunneling coupling between the wells. This model is only integrable for either $W=0$ (non-interacting case) or $J=0$ (no dynamics) as one may easily verify. If both energy scales $J$ and $WN_{\rm fill}$, where $N_{\rm fill}$ is the average filling, are comparable, the Bose-Hubbard model is a paradigm for a quantum chaotic system \cite{Kolo2004}. The Hamiltonian above well describes an ultracold atomic gas in sufficiently deep lattices where the ground band is very flat and decoupled from higher lying energy bands of the periodic lattice \cite{BDZ2008}. 

Many extensions of the basic model of Eq. (\ref{eq:1}) have been recently studied \cite{BDZ2008, EBHM, Kordas2015}. To be more specific, models with more than the usual single-band approximation were investigated in references \cite{Morebands1, Morebands2, Morebands3, Morebands4}, additional gravitational forces in \cite{Kolo2003, TMW2007}, or non-local interactions in \cite{Bloch2008, Pfau2009, Longrange, AW2012, KLW2012}. Our goal is here to present a model including two coupled energy bands. Such a system is implemented in ongoing experiments, for instance at Innsbruck \cite{Innsbruck2013, Innsbruck2014, Innsbruck2014a}, Harvard \cite{Greiner2011a, Greiner2011b} and Munich \cite{Bloch2008, Bloch2011}, and finds application in other realizations of Wannier-Stark lattice systems with ultracold bosons \cite{Sias2007, Zene2008}. Because of the non-local interband couplings, our minimal two-band model has similar quantum chaotic properties as its simplest version (\ref{eq:1}), see also refs. \cite{PMW2013,PhDThesis2013}. We will show now how these properties can be used to steer many-body Landau-Zener dynamics \cite{Oka2010} and to investigate the quantum (thermo)dynamics of the isolated many-body quantum system. 

\section{Minimal two band model}\label{sec:2}

Our two band model, significantly extending the Hamiltonian of Eq. (\ref{eq:1}), for locally interacting bosons is given by the following Hamiltonian
\begin{eqnarray}\label{eq:2}
 \hat{H}_{\rm 2B} &=&\sum_{l,\beta}\left[-\frac{J_{\beta}}{2}\left(\hat{\beta}^{\dagger}_{l+1}\hat{\beta}_{l}+h.c.\right)+
 \frac{W_{\beta}}{2}\hat{\beta}^{\dagger 2}_{l}\hat{\beta}^2_{l} + 2\pi F l \hat n_l^{(\beta)}\right]\,\nonumber\\
 &+& 2\pi F \sum_{l,\mu} \left[ C_{\mu} \hat{a}^{\dagger}_{l+\mu}\hat{b}_l + h.c. \right]+
\frac{W_x}{2} \sum_l \left[ \hat{b}^{\dagger 2}_{l}\hat{a}^2_{l}+h.c. \right] \nonumber\\
 &+& 2W_x \sum_l \hat{n}^{(a)}_l\hat{n}^{(b)}_l+\frac{\Delta}{2} \sum_l (\hat{n}^{(b)}_l-\hat{n}^{(a)}_l) \,.
\end{eqnarray}
This model preserves the total number $N$ of bosons on $L$ lattice sites. Then the full dimension of the state space (Fock space) is given by
\begin{equation}\label{eq:3}
D=\frac{(N+2L-1)!}{N!(2L-1)!}\,.
\end{equation}

\begin{figure}
  \includegraphics[width=\columnwidth]{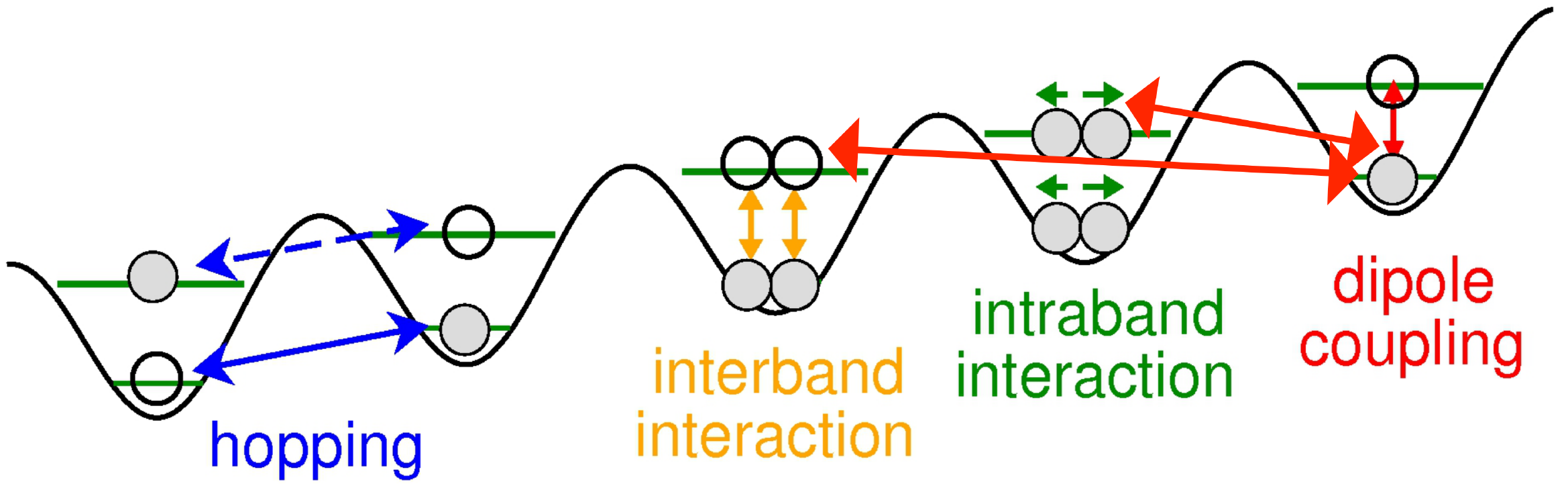}%
  \caption{\label{fig:1}
    Sketch of the system we model by the Hamiltonian $(\ref{eq:2})$. The hopping terms represent $J_a$ (solid blue arrow) and $J_b$ (dashed blue arrow). Intraband and  interband interactions correspond to the terms with $W_{a,b},W_x$, respectively. The field induced dipole couplings are shown by the red arrows for  $C_{\mu}$, with $\mu=-2, -1, 0$. All the latter turn out to be relevant at the corresponding resonant tunneling conditions as the next two figures report.
    }
\end{figure}

The operators $\hat\beta_l$ and $\hat\beta^{\dagger}_l$ represent the annihilation and creation operators. $\hat n^{(\beta)}_l=\hat\beta^{\dagger}_l\hat\beta_l$ is the number operator, with band index $\beta \in \{a,b\}$ for the lower and the upper band, respectively. The parameter space is defined by the parameters $(J_{\beta},W_{\beta},W_x,C_{\mu},\Delta)$. Most of the terms are sketched schematically in Fig. \ref{fig:1}. $J_{\beta}$ is the kinetic energy scale, characterizing the nearest neighbor hopping matrix elements. The $W$'s represent the on-site, intra- and inter-band interaction strengths. $2\mu+1$ dipole coupling coefficients $C_{\mu}$ are induced by the constant gravity or Stark field $F$, with $\mu\in \mathbb{Z}$. The Stark force itself is given by the third term in the first line of Eq.~(\ref{eq:2}). The average energy band gap is given by $\Delta$. Two-body interactions between the local onsite single-particle energy band levels lead to the interband exchange term in the second line of Eq.~(\ref{eq:2}).

In contrast to previous versions of the above model \cite{PMW2010,TMW2007}, we explicitly include here more cross-band couplings with $|\mu|\leq 2$, for which $\mu=\{-2,-1,0,1,2\}$. At least for resonant tunneling restricted to the next two neighboring sites, more terms are not necessary since the corresponding coupling coefficients go fast to zero with increasing $|\mu|$ \cite{PhDThesis2013}. Yet, as we shall see non-onsite cross terms are very relevant to describe, for instance, resonant tunneling between the two energy bands induced by the Stark field. Resonant tunneling finds important applications in solid-state devices \cite{Chan74} and is studied experimentally with ultracold atoms in the mean-field \cite{Sias2007,Zene2008} and many-particle regime \cite{Innsbruck2013, Innsbruck2014, Innsbruck2014a}. In our sketch of Fig. \ref{fig:1}, resonant tunneling is almost realized between the upper level of the middle well and the lower level of the rightmost site. The resonance conditions are given for specific values of the Stark force, which is well controllable experimentally \cite{Innsbruck2013, Innsbruck2014, Innsbruck2014a, Sias2007, Zene2008}. From Fig. \ref{fig:1} it should be clear that the addition of the coupling terms $C_{\pm 1}$ and $C_{\pm 2}$ is crucial for investigating, in particular, first and second-order resonant tunneling, i.e. resonant tunneling to the nearest or next nearest neighbor well, respectively.

The non-integrability of Hamiltonian (\ref{eq:2}) makes necessary a numerical treatment based on exact diagonalization or explicit integration schemes. For the former, we actually transform the static Stark terms into periodically time-dependent phase factors for the hopping parameters and diagonalize the corresponding time-independent Floquet matrix. The temporal evolution for non-periodic time-dependence is done with a highly optimized Runge-Kutta scheme taking into account only the non-zero elements for matrix-vector multiplications. More details on numerical procedures are discussed elsewhere \cite{PMW2014b}. Here we focus on the physical consequences of the strong coupling in dynamical simulations. In this regime, the actual finite size of the system expressed in number of atoms $N$ and lattice sites $L$ is not really crucial since the total size $D$ of the accessible Hilbert space is anyway large for $N,L \geq 4$ \cite{PhDThesis2013, PMW2013, PMW2014b}. Hence we can restrict to rather small $L=4$ and $L=5$ with filling of order one for numerically expensive time-dependent computations.

\begin{figure}
  \includegraphics[width=\columnwidth]{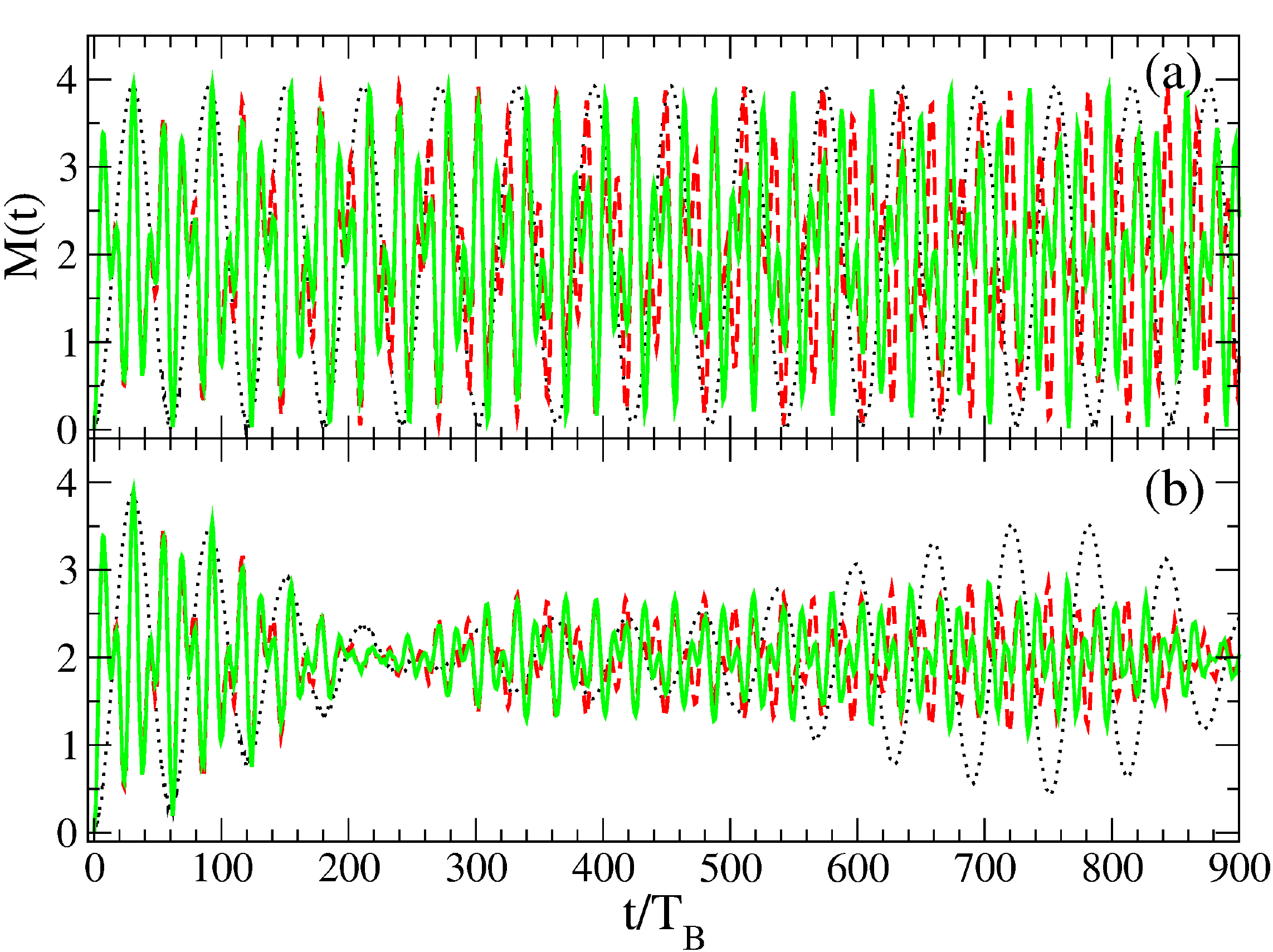}%
  \caption{\label{fig:2}
    The temporal evolution of the initial state $\ket{\psi_0}=|1111; 0000 \rangle$ ($N=4=L$) in time units of the Bloch period $T_B=1/F$. 
    Shown are the upper band populations $M(t)$ for (a) $W_{a,b,x}=0$ and (b) weak interactions 
    $W_a=2.1\times 10^{-3}, W_b=2.6\times 10^{-3}, W_x=2.3\times 10^{-3}$. 
    The different curves are computed including only
    $C_0=-0.09$ (black dotted lines), $C_0$ and $|C_{\pm 1}|=0.039$ (red dashed lines), and finally the full model dynamics with all coupling elements 
    $C_{0, \pm 1, \pm 2}$ with $|C_{\pm 2}|=2.1\times 10^{-3}$ (green solid lines). The almost perfect oscillations in (a) and the revival in (b) around $t/T_B = 750$
    are heavily affected by the additional coupling terms. In particular, the phase of the interband oscillations depends on the specific model and the 
    revival in (b) disappears in the background when including $C_{\pm 1, \pm 2}$. The remaining parameters are $F=0.19, \Delta=1.16 , J_a=0.08 , J_b=-0.12$.
    }
\end{figure}

\begin{figure}
  \includegraphics[width=\columnwidth]{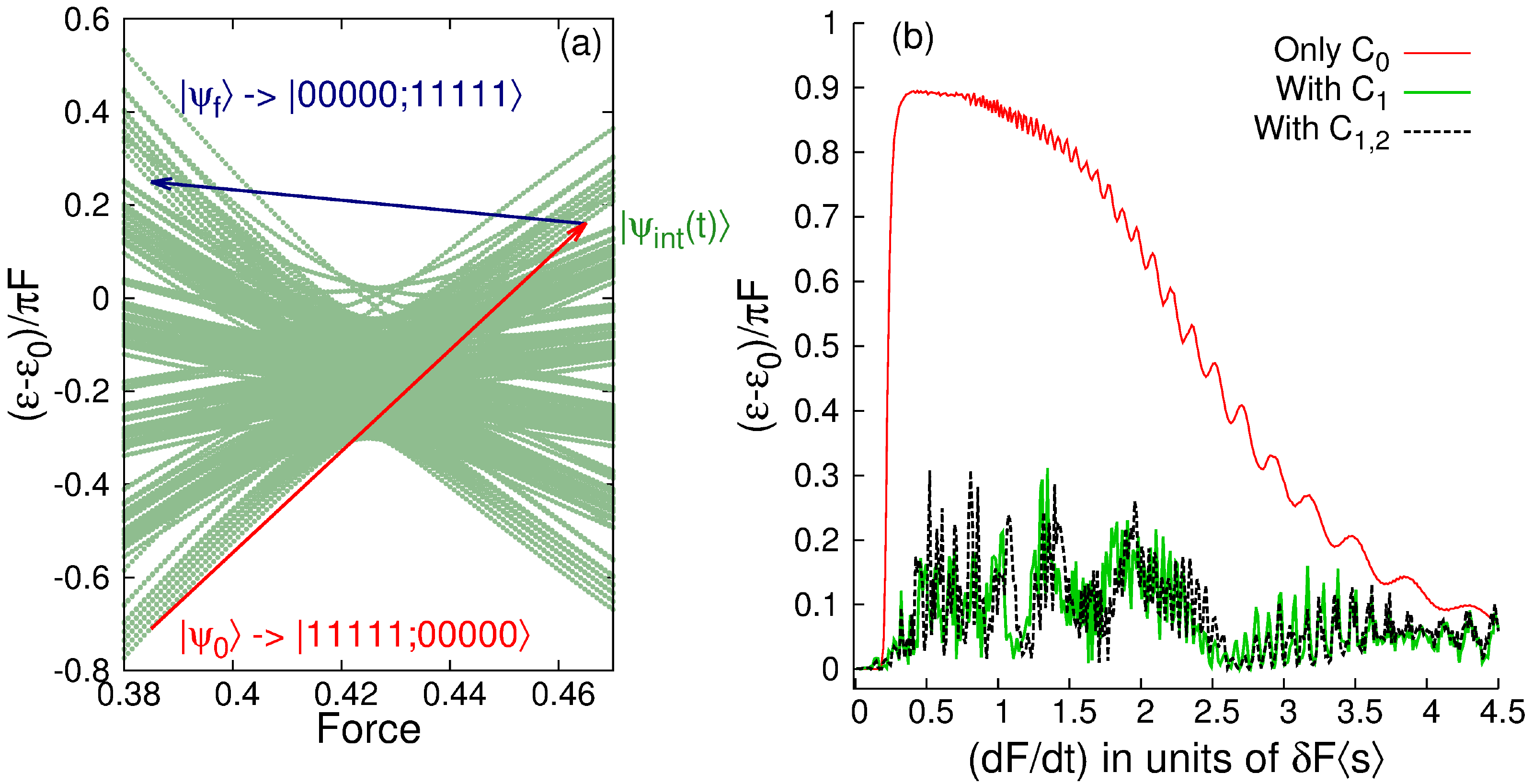}%
  \caption{\label{fig:3}
    Dynamical numerical experiment to prove the importance of field inter band cross terms $C_{\mu}$ with $\mu \neq 0$. (a) Typical many-body energy spectrum across resonant tunneling characterized by a single-particle avoided crossing (filled by dense lying many-body levels). We try to transfer an initial state $\ket{\psi_0}$ via a fast (diabetic) evolution to an intermediate state $\ket{\psi_{\rm int}}$ via a sweep with the rates shown on the x axis in (b) into the final state $\ket{\psi_{\rm f}}$. The arrows indicate the start of the protocol at $F_0 \approx 0.385$ over the intermediate value $F_{\rm int} \approx 0.465$ back to $F_{\rm f}=F_0$.(b) The success probability of transferring a large population from a Mott-like lower band state into a corresponding upper band state as a function of the slope of the force sweep (in units of $\delta F \langle s \rangle$ as defined in the main text).  In the case of $C_\mu=0$ for $\mu \neq 0$ the success is high (red/upper solid line). The inclusion of more relevant couplings $C_{\pm 1}$ (green/lower solid line) and both $C_{\pm 1, \pm 2}$ (black dashed line) destroys this possibility. While the details depend on the specific model used for the computation, we observe that the crucial additional term is $C_{\pm 1}$ for the first-order resonance studied here for $L=5=N$. Our realistic parameters are \cite{PhDThesis2013}: $\Delta=0.32$, $J_a=0.041$, $J_b=-0.046$, $W_a=0.027$, $W_b=0.029$, $W_x=0.028$, $C_0=-0.096$, $|C_{\pm 1}|=0.046$, $|C_{\pm 2}|=0.007$.}
\end{figure}

The importance of including the interband non-local couplings $C_{\pm1, \pm2}$ close to resonant tunneling conditions is highlighted in Figs. \ref{fig:2} and \ref{fig:3}. 

First, we study the temporal evolution of the upper-band population for an initial state with filling one and all atoms in the ground band. The upper-band occupation number is given by the expectation value
\begin{equation}\label{eq:4}
M(t) = \sum_l \langle \phi (t) |n^{(b)}_l \ket{\phi (t)},
\end{equation} 
with $|\phi (t) \rangle$ an arbitrary Fock state. Figure \ref{fig:2} (a) shows this number for $N=4=L$ versus time for no interactions $W_{a,b,x}=0$. While for the case with only $C_0$ nonzero almost perfect interband oscillations occurs, the addition of the terms with $|\mu| > 0$ complicates the dynamics. Hence, the results are different even for the non-interacting problem. Here the simplified theory of ref. \cite{PMW2010} predicts an oscillation period which is independent of the system size, here in particular of $N$ and $L$ at constant $N/L=1$. Taking into account the terms $C_{\mu}$, with $\mu=\pm 1, \pm 2$, the oscillations show additional periods and the specific result depends on the system size $L$ instead (not shown). The situation is even more dramatic if we look at an interacting case in  Fig. \ref{fig:2} (b), for which the revivals, seen if only $C_0$ is considered (dotted line), are degraded once we include the $C_{\pm 1, \pm 2}$ coupling terms (see solid line for the full model for instance). This means that the periodic revivals seen for the non-interacting case and just $C_0$ couplings disappear not only because of stronger interactions, but also due to the additional couplings $C_{\pm 1, \pm 2}$. This implies that -- even in the non-interacting case -- the two-band system cannot be mapped any more onto a simple quantum spin Ising model, as done e.g. in \cite{Greiner2011a, PSW2011}.

Secondly, we investigate the temporal evolution of an eigenstate of the two band system which has again the form $\ket{\psi_0} \approx |11111; 00000 \rangle$ at a specific value of the force $F=F_0$. Then we sweep the force linearly in time $F(t)= {\dot F}t + F_0$ and follow the dynamics of the corresponding many-body state $\ket{\psi (t)}$. The goal of our optimization procedure is to realize a final state close to $ |00000; 11111 \rangle$, which would correspond to transferring a Mott-like state from the ground band to the second band. The following protocol turned out to be feasible, see panel (a) in Fig. \ref{fig:3}: 
\begin{itemize}
\item 
(i) the initial and the intermediate state are chosen around but slightly outside the resonant tunneling region of width $\delta F$ in the spectrum (see panel (a) in Fig. \ref{fig:3}); more precisely, the initial state on the left at $F_0$ and the intermediate state at $F_{\rm int}$ on the right at of this region. The final state is obtained at the same force value where we started from $F_{\rm f}=F_0$.
\item
(ii) we diabatically evolve, i.e. with large sweeping rate ${\dot F} \gg \delta F \langle s\rangle$, the initial condition from left to right. Here 
$\delta F$ is the width of the resonant tunneling region and $\langle s \rangle$ the mean level spacing. This procedure ensures that the intermediate state $\ket{\psi_{\rm int}}$ is of a similar form in the Fock basis as the initial one, yet its energy is increased. 
\item
(iii) Now we invert the evolution, i.e, the state $\ket{\psi_{\rm int}}$ is evolved backwards decreasing the force again. We compute the success probability, i.e., the probability of obtaining as a final state $\ket{\psi_{\rm f}}=|00000; 11111 \rangle$ as a function of the slope ${\dot F}$. 
\end{itemize}
In panel (b) of Fig. \ref{fig:3} we show the success probability (solid red line). The latter shows a clear stable maximum over a sufficiently broad range of slopes (plateau region), while for too large slopes Rabi-like oscillations occur. The plateau region is obtained for sweeping rates values which are close to but smaller than the product of the effective width $\delta F$ of the many avoided crossing and $\langle s \rangle$. Hence, if only the coefficient $C_0$ is included the protocol above would produce reliably the target state. Including higher order couplings, our protocol becomes unstable and useless for a production of this particular state with high fidelity. For nearest-neighbor resonant tunneling, the main contribution with respect to the naive model (red/upper solid line) arises from $C_{\pm 1}$, while the next order $C_{\pm 2}$ will become important for longer-range resonant tunneling along the lattice (not shown here, but studied in the mean-field regime experimentally in \cite{Sias2007, Zene2008}).

The two examples shown in Figs. \ref{fig:2} and \ref{fig:3} show that we cannot neglect non-local band exchange terms $C_{\mu}$, with $\mu=\pm 1, \pm 2$, in particular at resonant tunneling conditions \cite{Sias2007,Zene2008}. On the one hand, it denies the possibility to prepare interesting higher-band states, if not very sophisticated protocols for optimizing the transport across the interband avoided crossing region are found. On the other hand, only the inclusion of next-nearest neighbor cross couplings leads to a complex mixing of the states corresponding to the instantaneous spectra for fixed forces $F$, respectively. This chaotic level mixing and its consequences will now be characterized further in the next section.

\section{Quantum thermodynamics}\label{sec:3}

Since our model conserves the particle number $N$, the sum of the lower and upper band population obviously is $\sum_l \left(n^{(a)}_l+n^{(b)}_l \right)=N$. In the limit of vanishing interactions $W_{\beta},W_x \to 0$, the spectrum of $(\ref{eq:2})$ can be split into sets of states with the same upper-band occupation number \cite{PMW2013, PMW2014b}, see Eq.~(\ref{eq:4}). These sets of states, we call them $M$ manifolds in the following, are very useful also for the characterization of the complex spectra of interacting systems. As shown in detail in ref. \cite{PMW2013}, close to resonant interband tunneling, these manifolds lose their property of being good quantum numbers due to strong interactions. This indicates that the spectra are strongly mixed due to the interactions {\em and} the avoided crossing, which squeezes the levels close together (see Fig. \ref{fig:3} (a)). 

Other useful quantities for the characterization of strong level mixing are 
\begin{equation}\label{eq:5}
\theta_{\beta=a,b} = \left\langle\frac{1}{2}\sum_l \hat{\beta}^{\dagger 2}_{l}\hat{\beta}^2_{l} \right\rangle_{\phi}
\end{equation}
 and 
\begin{equation}\label{eq:6}
\theta_x=2 \left\langle \sum_ln^{(a)}_ln^{(b)}_l \right\rangle_{\phi}.
\end{equation}
They represent the onsite intraband particle interactions and the onsite interband interactions, respectively. Obviously, both quantities are zero in the interaction-free case.

\begin{figure}
  \includegraphics[width=\columnwidth]{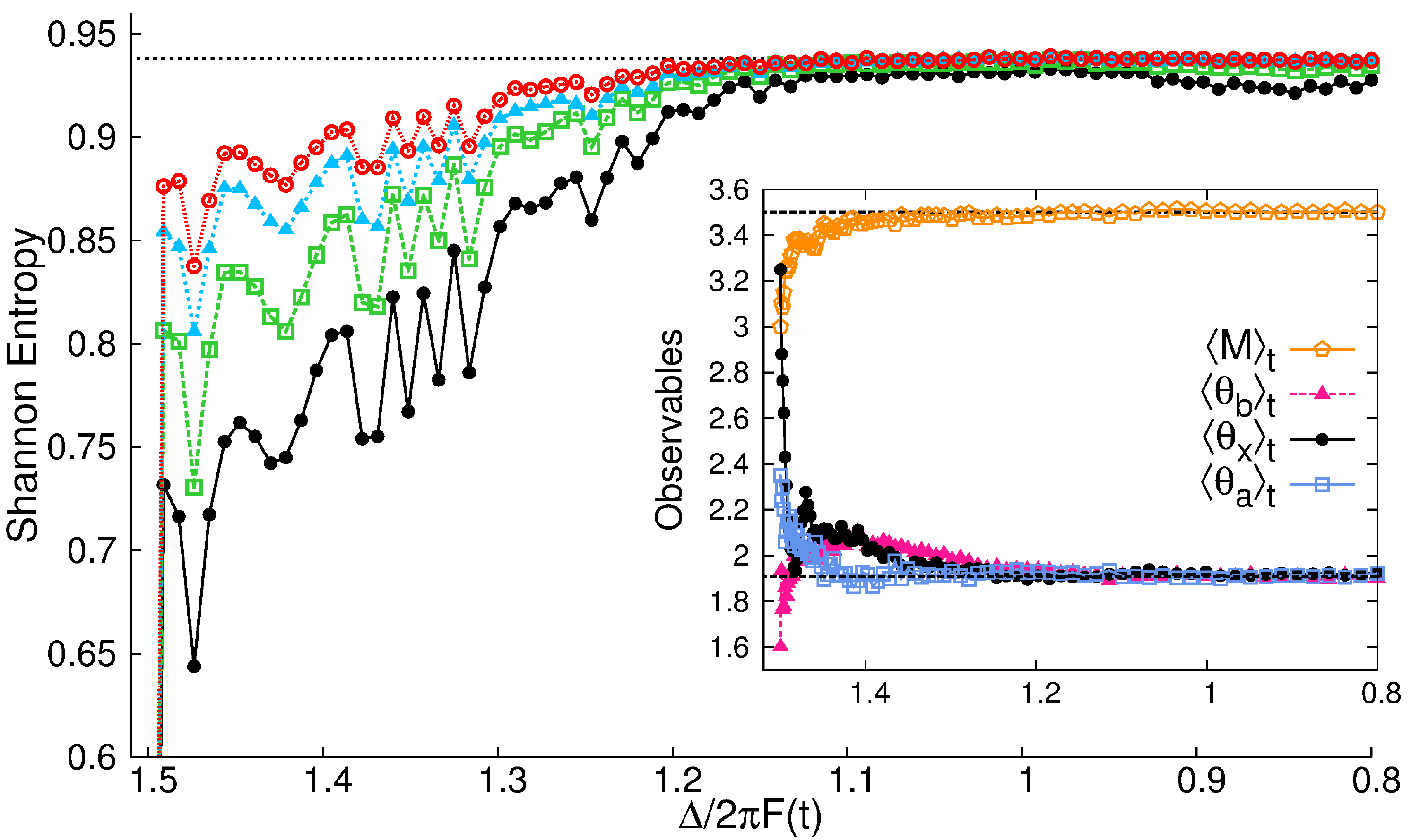}%
  \caption{\label{fig:4}
  Shannon entropy in the instantaneous basis, see Eq. (\ref{eq:8}), for $N=6, L=5$. The different colors represent different M manifolds of the initial state: $M = 0$ (black line with circles), $M = 1$ (green dashed line with squares) and $M = 2$ (light/blue dashed line with triangles) and $M = 3$ (red dashed line with open circles). The inset shows the
  temporal evolution of the single and two-body observables $\{\hat M,\hat \theta_{\beta,x}\}$. All of them relax quickly to their equilibrium value given in Eq. (\ref{eq:9}) and plotted as constant black dashed lines. The other parameters are identical to the ones of Fig. \ref{fig:3} and correspond for our filling $N/L=6/5$ to a regime of strong level mixing due to interparticle interactions.
    }
\end{figure}

We want to study now better the non-equilibrium dynamics in the presence of interactions comparable to the hopping strength. As in the previous section, we drive an initial state across the resonant tunneling regime, where many non-adiabatic transitions take place. For this, we use a linear sweeping function, which is indeed inspired by Landau-Zener transition models for our system \cite{Loerch2010}: $F(t)= {\dot F}t + F_0, \, t > 0$. The rate ${\dot F}$ is chosen in such a way to be similar to the product of the mean level spacing $\langle s\rangle$ and the total width $\delta F$ in the parameter $F$ of the strongly coupled region in the spectrum. This guarantees an optimal sweeping rate for strong level coupling in the dynamics, avoiding too diabatic (direct crossing of the region without level spreading) or too adiabatic (following essentially one level only) evolutions. The scan starts slightly before the avoided crossing at resonant tunneling and stops slightly afterwards. The presence of avoided crossings in the spectrum, see Fig.~\ref{fig:3} (a), generates then a spreading of the initially localized wave packet in the instantaneous eigenbasis of states $\ket{\epsilon_i}$ with eigenenergies $\epsilon_i(F(t))$. The local density of states (LDOS), 
defined by 
\begin{equation}\label{eq:7}
P_{\psi}(\varepsilon,g)=\sum_i|C_i|^2\delta(\varepsilon-\varepsilon_i),
\end{equation}
with $C_i\equiv \langle \psi_t|\varepsilon_i\rangle$, characterizes 
this spreading. For strong interactions, a nearly flat distribution is reached. Here $|C_i|^2\sim 1/D$, i.e. the system obeys a equipartition condition. The spreading may also be analyzed using the Shannon information entropy \cite{Smilansky1987} 
\begin{equation}\label{eq:8}
S_{\rm sh}=-\sum_i|C_i|^2\ln |C_i|^2,
\end{equation}
which approaches $S_{\rm  sh}\approx\ln D$ in statistical equilibrium. The evolution of the entropy is shown for typical system parameters in Fig.~\ref{fig:4}. The entropy starts out from a small value, which is zero when we start exactly with an eigenstate of the system for $F=F_0$. Then it systematically increases saturating after crossing the minimum of the avoided crossing. The saturation values are all identical and close to one (actually $0.94$ as predicted by random matrix theory for full chaotic level mixing \cite{Zele1996, Fausto2012}).

The thermalization of observables in a complex quantum system can be investigated with the help of the eigenstate thermalization hypothesis \cite{Deutsch1991, Srednicki1994, Therm}. For this, we check whether the expectation value of our operator $\hat O$ of interest and its time average approach the diagonal approximation:
\begin{equation}\label{eq:9}
\langle\hat O\rangle_t \equiv \langle\psi_t|\hat O|\psi_t\rangle ,~ \overline{O} \equiv \lim_{t\rightarrow\infty}\frac{1}{t}\int^{t}_0\langle\hat O\rangle_{t'}dt'\longrightarrow \sum_i|C_i|^2O_{ii}\,,
\end{equation}
where $O_{ij}=\langle \varepsilon_i|\hat O|\varepsilon_j\rangle$. Since we are treating a closed system, the density matrix is represented by the pure state $|\psi_t\rangle$ for the full system. In the regime of strong chaotic level mixing, the distribution of the coefficients $|C_i|$ after the sweep is essentially flat confirming the statistical relaxation in our isolated system (see the discussion of the Shannon entropy above). This regime can indeed be characterized by an effective temperature of the size of the spectral width divided by the Boltzmann constant, see ref. \cite{PMW2014a} for the somewhat subtle definition of temperature in our isolated system of interacting particles in terms of the instantaneous Floquet bases at fixed forces $F$.

The coherent dephasing arises from passing the zone with many avoided crossings with a broad distribution of widths as expected from random matrix theory, see e.g. \cite{widths}. The passing of this chaotic zone close to resonant tunneling conditions makes the off-diagonal elements $O_{ij}$ go to zero quickly and the evolution can be well approximated by the diagonal contributions only. We study the time evolution of the set of observables $\{\hat M,\hat \theta_{\beta,x}\}$ introduced above. Optimal thermalization in the instantaneous eigenbases is obtained for a sweeping parameter $\dot F/\delta F \langle s\rangle$ of order 1. This condition corresponds to the strong mixing condition mentioned already in the previous section. All the expectation values shown in the inset of Fig.~\ref{fig:4} converge to their microcanonical averages (in our case identical to the diagonal ensemble on the right hand side of Eq. (\ref{eq:9})) via quantum chaotic diffusion across the instantaneous spectrum. Hence, the right hand side of Eq. (\ref{eq:9}) is, in practice, not any more dependent on the initial state because of the strong level mixing and the resulting irreversibility of the quench dynamics (see \cite{PMW2014a} for details).

\section{Conclusions and perspectives}\label{sec:4}

We presented a model of two energy bands coupled by internal particle-particle interactions and by an external Stark force. 
Our model may be realized with ultracold bosons in periodic optical lattices \cite{Innsbruck2013, Innsbruck2014, Innsbruck2014a}. The non-intergrability of our quantum many-body system can be used to engineer complex dynamics. The temporal evolution of the system across a region of strong level clustering (avoided crossings) highlights the relevance of additional interband coupling terms. Furthermore, it allows us to investigate the chaotic diffusion and non-equilibrium properties of single and two-body observables within the instantaneous quantum spectra. 
An advantage of our system is that the sweep across the chaotic spectral region at resonant tunneling conditions can be done rather fast (also because the region is small in parameter space, see e.f. Fig. \ref{fig:3}(a)), and one does not have to wait asymptotically long for reaching thermalization. Together with possible experimental implementations with ultracold atoms, our results shed light on the understanding of thermalization in closed non-integrable systems through the eigenstate thermalization hypothesis \cite{Deutsch1991, Srednicki1994, Therm}. Other applications include the study of reversibility properties \cite{PMW2014a, Richter2014} and the deterministic production of entanglement \cite{Luba2011} in driven many-body quantum systems.
 
\section{\bf Acknowledgments}
We kindly acknowledge support by the DFG  (grant number WI 3426/7) and by the COST Action MP1209 ``Thermodynamics in the quantum regime". Furthermore, we thank Thomas Wellens for stimulating the study of more complex interband coupling models.


\end{document}